# Automated Experiments of Local Non-linear Behavior in Ferroelectric Materials

Yongtao Liu,[1] Kyle P. Kelley,[1] Rama K. Vasudevan,[1] Wanlin Zhu,[2] John Hayden,[2] Jon-Paul Maria,[2,3] Hiroshi Funakubo,[4] Maxim A. Ziatdinov,[1,5,a] Susan Trolier-McKinstry,[2,3] and Sergei V. Kalinin[6,b]

[1] Center for Nanophase Materials Sciences, Oak Ridge National Laboratory, Oak Ridge, TN 37831, USA

[2] Department of Materials Science and Engineering, The Pennsylvania State University, University Park, PA 16802, USA

[3] Center for Dielectrics and Piezoelectrics, Materials Research Institute, The Pennsylvania State University, University Park, PA 16802, USA

[4] Department of Material Science and Engineering, Tokyo Institute of Technology, Yokohama 226-8502, Japan

[5] Computational Sciences and Engineering Division, Oak Ridge National Laboratory, Oak Ridge, Tennessee 37831, USA

[6] Department of Materials Science and Engineering, University of Tennessee, Knoxville, TN 37916

We develop and implement an automated experiment in multimodal imaging to probe structural, chemical, and functional behaviors in complex materials and elucidate the dominant physical mechanisms that control device function. Here the emergence of non-linear electromechanical responses in piezoresponse force microscopy (PFM) is explored. Non-linear responses in PFM can originate from multiple mechanisms, including intrinsic material responses often controlled by domain structure, surface topography that affects the mechanical phenomena at the tip-surface junction, and, potentially, the presence of surface contaminants. Using an automated experiment to probe the origins of non-linear behavior in model ferroelectric lead titanate (PTO) and ferroelectric $Al_{0.93}B_{0.07}N$ films, it was found that PTO showed asymmetric nonlinear behavior across *a/c* domain walls and a broadened high nonlinear response region around *c/c* domain walls. In contrast, for $Al_{0.93}B_{0.07}N$, well-poled regions showed high linear piezoelectric responses paired with low non-linear responses and regions that were multidomain indicated low linear responses and high nonlinear responses. We show that formulating dissimilar exploration strategies in deep kernel learning as alternative hypotheses allows for establishing the preponderant physical mechanisms behind the non-linear behaviors, suggesting that this approach automated experiments can potentially discern between competing physical mechanisms. This technique can also be extended to electron, probe, and chemical imaging.

[a] ziatdinovma@ornl.gov
[b] sergei2@utk.edu

Ferroelectric materials exhibit a broad spectrum of non-linear behaviors in response to frequency-dependent electric and elastic stimuli. These behaviors are vital for applications of ferroelectric materials – as much as 75% of the observed piezoelectric response in lead zirconate titanate ceramics are associated with nonlinear extrinsic contributions.[1-11] These nonlinearities can both enable new device concepts and limit applicability of extant ones.[12] Non-linear behaviors are strongly tied to the physics of ferroelectric materials; intrinsic nonlinearities produce a field-dependence in properties such as the dielectric susceptibility, even in single domain single crystals.[11, 13, 14] Practically, however, extrinsic responses, typically associated with the motion of domain walls and phase boundaries, dominate macroscopic nonlinearities in ferroelectric materials, producing emergent behaviors from mesoscopic to macroscopic length scales.[15] As such, non-linear dynamics have been used to explore ceramics[16] and single crystals[13, 17, 18] and quantum phase transitions,[19] addressing long-standing issues in the physics of ferroelectric materials. Similarly, the coupling between piezoelectric and dielectric phenomena leads to intertwined non-linear responses.[20]

Recently, it has been shown that SPM can be used to detect the non-linear electromechanical responses in point spectroscopy modes.[21] This approach was extended to probe the ac bias dependence of piezoresponse via hyperspectral measurements and local non-Rayleigh responses.[22-24] Local Raleigh-like[25] responses have been used to explore the effect of single grain boundaries on wall motion,[26-28] and establish the local relationships between non-linearity and minor hysteresis loops.[29]

However, in local measurements, non-linear responses can originate from various phenomena, including intrinsic dielectric and piezoelectric tunability, domain wall dynamics, electrostatic interactions,[30, 31] and elastic non-linearities at the tip-surface junction[25]. These effects are coupled at the tip-surface junction and cannot be differentiated by the bias dependence of local PFM signal alone. Similarly, the contact area, elastic softening, and indentation modulus cannot be separated based solely on resonant frequency shift measurements, and hence changes in resonance frequency can stem both from variations in Young's modulus due to the elastic softening and changes in surface topography.[32-34] However, it can be argued that insight into these mechanisms can be obtained based on statistical data analyses of the spatial features in the data, building correlations between topography, resonance frequency shifts, polarization gradients, and detected non-linear responses.

Traditionally, such correlations have been built based on the analysis of spectroscopic data sets acquired on a dense square grid of points.[25] In this case, the high-resolution structural and functional images can be co-registered with the hyperspectral images containing bias dependence of the response. This approach is time consuming. More importantly, it necessitates sampling all the correlations present in the data set *equally* and the small number of the regions with interesting behaviors (e.g. large or small responses) can remain unnoticed. As a result, identification of the relevant physical mechanisms from the correlative data is hindered.

An alternative approach is that of the automated experiment, uncovering the correlations between local structural and functional properties following a specific reward (e.g. curiosity or discovery of specific behaviors). Previously, this approach was used for automated PFM experiments based on a combination of deep kernel learning (DKL) and a physics-based reward function.[35] In that work, a single shot DKL model learned the correlation between the structural descriptor (e.g., a patch of topographic or PFM amplitude image) and the local materials response (hysteresis loop). The exploration of the image plane was controlled by the reward function (scalarizer), that specified which aspect of the predicted response is of interest. As a result, it was possible to explore which element of the domain structure is associated with large hysteresis loop areas in the on-field and off-field regime.

Here, this concept is extended to explore the role of multiple possible channels in evolution of piezoelectric nonlinearity in PbTiO$_3$ (PTO) and Al$_{0.93}$B$_{0.07}$N. Deep kernel learning is applied to band excitation amplitude spectroscopy (BEAM) to explore the correlation of nonlinearity and different band excitation PFM (BEPFM) image channels, i.e. topography, amplitude, phase, and resonance frequency. PTO and Al$_{0.93}$B$_{0.07}$N ferroelectric films were used as model systems with very different levels of piezoelectric nonlinearity.

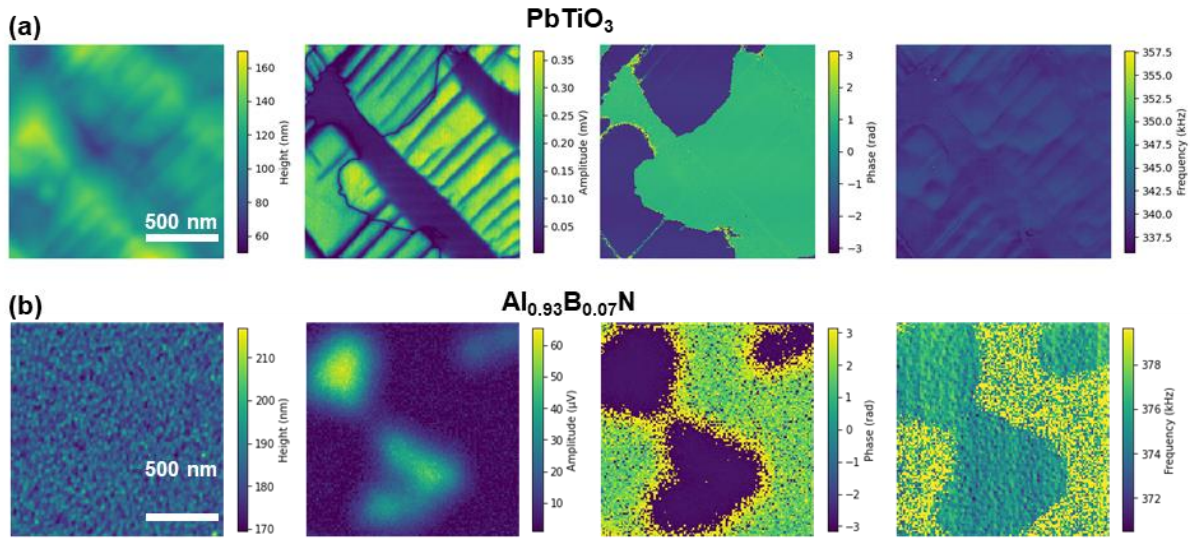

**Figure 1.** BEPFM topography, amplitude, phase, and resonance frequency results of **(a)** PbTiO$_3$ thin film and **(b)** Pt/Al$_{0.93}$B$_{0.07}$N/W capacitor.

A PTO thin film was grown via metalorganic chemical vapor deposition (MOCVD) with a SrRuO$_3$ (SRO) conducting buffer layer on (001) KTaO$_3$ (KTO) substrates, as reported by Morioka et al.[36] Previous studies on this film show (1) the causal relationships between various channels of the BEPFM data[37], (2) the correlation between local domain structure and properties encoded in the polarization-electric field hysteresis[38]. Figure 1a shows the BEPFM data of this PTO thin film, indicating both out-of-plane *c* domains and in-plane *a* domains. In addition, the ferroelastic domains are visible in topography images due to the different surface inclinations between adjacent domains. As expected, variations in the elastic stiffness between adjacent domains and topographic variations produced a rich structure in the resonance frequency shift image.

A second exemplar was a 205 nm thick Pt/Al$_{0.93}$B$_{0.07}$N/W capacitor on a sapphire substrate. The Al$_{0.93}$B$_{0.07}$N film preparation is described in the Methods section. In this case, the field across the device structure is uniform, and the probe acts as a sensor of mechanical displacement induced by the piezoelectric effect. This sample was originally woken up, and left in a state in which some regions were left largely unipolar, and other regions had both up and down domains throughout the film thickness. Figure 1b shows the BEPFM results of this Pt/Al$_{0.93}$B$_{0.07}$N/W capacitor. Amplitude, phase, and frequency images show domains with a lateral scale of hundreds of nanometers with variable piezoresponses. Imaging, domain writing, hysteresis loop measurements, and electric field cycling on macroscopic device[39-45] were recorded.

Following baseline characterization, the non-linear electromechanical behavior in these materials was studied. The local electromechanical response at each spatial grid location was measured via band excitation[46-48] as a function of ac bias.[49-51] Typically, the ac bias amplitude was swept following a triangular waveform; data from both upwards and downward branches were recorded. The data set represents the piezoresponse signal (PR) as a function of $V_{ac}$ amplitude and frequency $w$ at each spatial location ($x$, $y$), PR ($V_{ac}$, $w$, $x$, $y$). In classical BE analysis, the frequency dependence of the signal is analyzed as a harmonic oscillator to yield the amplitude, resonance frequency, $w_r$, Q-factor, and phase, $\phi$. The data set is analyzed using multivariate methods to study components and loading maps,[52-55] or disentangled representation via different autoencoders.[56-59] Similarly, correlative structure-property relationships, e.g. the correlation between domain structure and polarization dynamics, were established via encoder-decoder models and dual autoencoders.[60, 61]

Alternatively, the local spectral component can be fitted to a chosen functional form if the physical mechanisms behind the responses are known or hypothesized. For non-linearity measurements, the natural functional form is the polynomial expansion:

$$A\ (V_{ac}) = aV_{ac}^3 + bV_{ac}^2 + cV_{ac} \qquad \text{Eq (1)}$$

where $A\ (V_{ac})$ is the detected response amplitude and $V_{ac}$ is the excitation ac voltage. In this expansion, the offset term *a* corresponds to the intrinsic noise level in the system, *b* is the linear piezoelectric response, and higher-order terms *c* and *d* are non-linearities. The mechanisms behind bulk non-linear responses in ferroelectrics were explored by Damjanovic,[5, 6, 62] Mokry[63, 64] and others.[8, 25, 65, 66] As discussed above, the nonlinearity can be intrinsic; however, in most cases, the nonlinearities in ferroelectrics arise predominantly from motion of domain walls and/or phase boundaries.

When the measurements are performed locally via PFM, these responses can, in principle, be correlated with individual domain structure elements. However, in PFM, additional non-linearities emerge due to the responses at the tip-surface junctions and non-linear cantilever dynamics. They are exacerbated by surface topography and contaminates that affect local mechanical properties and electrical contact conditions. Previously, the origins of non-linear behaviors were explored using post-acquisition data analysis, as reported in Ref [25], where the data consists of a high-resolution microstructural image and the hyperspectral images of bias-dependence of the response. In addition to the time-consuming data acquisition process, this approach is limited to a few regions of interest, the sampling limitations of the rectangular grids, and by measurement latencies, which collects statistical information from known objects of interest, rather than uncovering structure-property relationships. More importantly, this approach equally samples all the correlations present in the data and the regions with interesting behaviors can remain unnoticed; consequently, the relevant physical mechanisms can potentially be missed. Alternatively, Kelley,[67] Liu,[68] and Volpe[69] have shown that in combination, computer vision and automated experiments can select locations for in-depth spectroscopic studies based on a-priori criteria in SPM. However, this approach relies on the a priori known objects of interest, rather then discovery of the microstructural elements having specific behaviors.

Previously, automated experiment (AE) workflows based on deep kernel learning (DKL) were implemented on the microscope [38, 70, 71] In DKL-AE, the spectroscopic measurements are first performed at several random locations, and the DKL builds correlative relationships between the structure images taken around measured locations and the corresponding spectroscopic responses; the predicted responses and uncertainties are used to determine the acquisition function for Bayesian optimization (or active learning); the largest measurement defines the next spectroscopic measurement location. DKL's acquisition function can either minimize uncertainty (pure exploration) or follow certain criteria (exploitation). For example, this strategy was used to discover the edge plasmons in EELS[70] and 4D STEM[86] and to establish the relationship between domain structure and minor hysteresis loops in PFM.[38]

Here the goal is to explore whether the mechanism behind the nonlinear electromechanical response can be discovered via DKL-AE in multimodal imaging by using predictability[72, 73] to identify physical mechanisms via active learning. Here the relationships (or lack thereof) between piezoelectric non-linearity and individual data channels including topography, resonance frequency shift, and the magnitude of the piezoresponse signal were specifically targeted. Accurate predictions[72] of functional responses correlates a material's function with the selected imaging channel. If DKL fails to establish a relationship between the selected channel and response of interest, the two are unrelated. However, if a particular channel has strongest predictive power for a specific parameter, this suggest that causal link between the two. Here this approach was demonstrated using pre-acquired data and implemented in the active learning framework during the automated experiment.

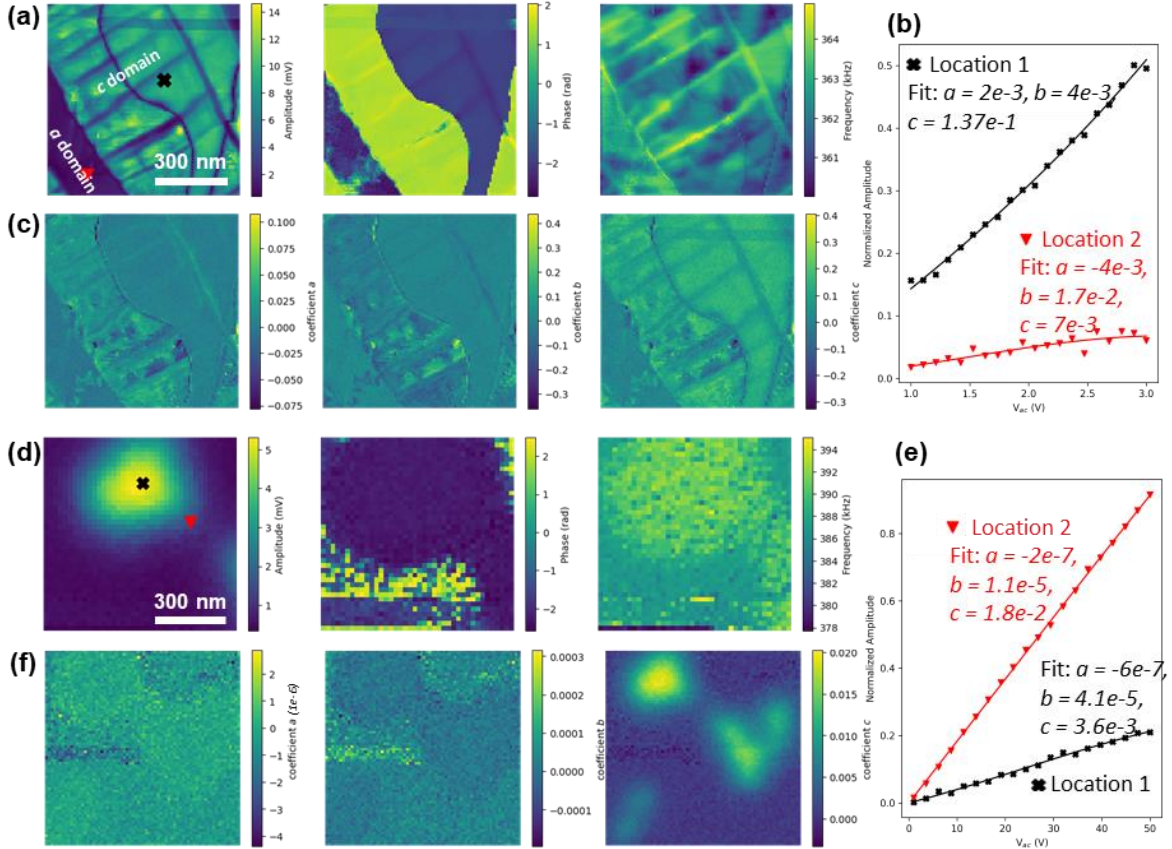

**Figure 2. BEAM results** of (**a-c**) PTO/SRO/KTO sample and (**d-f**) Pt/Al$_{0.93}$B$_{0.07}$N/W capacitor. (**a**), (**d**) averaged amplitude, phase, and frequency maps over all V$_{ac}$; (**b**), (**e**) two examples of piezoresponse amplitude *vs* V$_{ac}$ data with fitted curves from the location labeled on (**a**) and (**d**), respectively; (**c**), (**f**) three parameter (coefficient *a*, *b*, and *c*) maps resulted from fitting. The unit of coefficient *a, b,* and *c* are nm/V, nm/V$^2$, and nm/V$^3$, respectively.

To illustrate the principle of the DKL-AE, DKL was implemented using pre-acquired high density BEAM imaging datasets of the PTO thin film and Pt/Al$_{0.93}$B$_{0.07}$N/W capacitor, which provided the "known" ground truth image. Here, the piezoresponse resulting from applying a probing AC voltage (where the AC bias range depends on the material investigated) through the band-excitation method is acquired within a pre-defined grid, thus at each individual point a spectroscopic data of piezoresponse amplitude *vs.* AC voltage is available. Figures 2a-c show a 100*100 grid of BEAM data generated from the PTO film with bipolar AC excitation amplitude (~350-400 kHz) ranging from 1-3 V (well below the coercive voltage of 10V). To differentiate the effects of material nonlinearity and tip-surface junction nonlinearity, it is noted that the two should have different dependence on the frequency sweep direction. While the material nonlinearity should be independent of sweep direction, nonlinear interactions in the junction will appear as hysteresis in the response-curve even for the simplest cubic nonlinearities. In this work, the phase offset in the BEAM measurement was adjusted to minimize the tip-surface junction nonlinearities by minimizing the hysteresis as a function of the frequency sweep direction. BEAM measurements of the PTO and Pt/Al$_{0.93}$B$_{0.07}$N/W capacitors were carried out in ambient conditions.

Figure 2a shows the amplitude and frequency images across all AC voltages at each grid. Figure 2b shows two examples of piezoresponse amplitude *vs* AC voltage curves from the locations shown in Figure 2a. For this film, the displacement as a function of voltage curve was concave up. These curves were fitted to Equation 1 to produce three parameters: the cubic (*a*), quadratic (*b*), and linear (*c*) coefficients. Examplar fits and fitting parameters are shown in Figure 2b. Curve 1, from the *c*-domain, shows a larger nonlinear response (linear coefficient is $1.37 \times 10^{-1}$ nm/V) than curve 2 (linear coefficient is $7 \times 10^{-3}$ nm/V), from the *a*-domain. The $A$ ($V_{ac}$) at each grid was fitted using equation (2) in order to map the three parameters; this allows the analysis of the relationship between domain structures and $A$ ($V_{ac}$) of the system. Notably, grid points for in-plane *a*-domains have low out-of-plane piezoresponses, producing low or scattered signals in BEAM measurements. Figure 2c shows the maps of three parameters *a*, *b*, and *c*. As expected, the out-of-plane *c*-domains have a larger linear component with a linear coefficient around 0.4 nm/V, in contrast, the in-plane a-domains have a very small linear coefficient ~0 nm/V. The *c/c* domain walls have relatively smaller linear (~0 nm/V) and cubic components (~0 nm/V) but larger quadratic components (~0.4 nm/V$^2$). The *a/c* domain walls have a relatively larger cubic component (~0.1 nm/V$^3$) but smaller linear (~0 nm/V) and quadratic components (~0 nm/V$^2$).

Figure 2d-f shows the BEAM results of a Pt/Al$_{0.93}$B$_{0.07}$N/W capacitor. As described above, the sample was woken up into the ferroelectric state; some regions were left largely unipolar, and other regions had both up and down domains throughout the film thickness. This BEAM data is fitted as equation 1 to obtain three parameter maps. The linear *c* map appears to have a weak correlation between amplitude and frequency. It is likely that the high linear piezoresponse regions correspond to the unipolar regions in the sample.

With the ground truth behavior established, DKL was employed to elucidate structure-property relationships between the structural images (amplitude, phase, and frequency) and nonlinearity maps. DKL represents a hybrid of deep neural networks and Gaussian processes.[74] The neural network *g* parametrized by weights **w** embeds the high-dimensional structural image data *X* into the latent space where a standard GP kernel $k_{base}$ with hyperparameters **θ** operates. The corresponding regression model for a response variable *y* (here, the fit parameters in Eq 1) can be written using a standard Gaussian process formulation as:

$$y \sim MultivariateNormal(0, K_{DKL}(x, x')) \qquad \text{Eq (3a)}$$

$$K_{DKL}(x, x'|\boldsymbol{w}, \boldsymbol{\theta}) = k_{base}(g(x|\boldsymbol{w}), g(x'|\boldsymbol{w})|\boldsymbol{\theta}) \qquad \text{Eq (3b)}$$

The $K_{DKL}$ is referred to as a 'deep kernel' whose parameters (neural network weights and base kernel hyperparameters) are optimized with respect to the marginal likelihood resulting in an end-to-end learning scheme. Hence, the learned latent representation encodes the structure-property relationships and is different from standard dimensionality reduction techniques (such as principal component analysis or non-negative matrix factorization) where the 'reduction' of the high-dimensional image data is not conditioned on any observed physical functionality.

At the prediction stage, the trained DKL model produces the expected function value and associated uncertainty at new inputs $X^*$. These can be combined into the standard acquisition function used in active learning and Bayesian optimization to derive the next measurement point.

DKL was further implemented on the pre-acquired amplitude spectroscopy data set (i.e. ground truth data) of a model PTO film and a Pt/Al$_{0.93}$B$_{0.07}$N/W capacitor sample. Figure S1 and S2 shows the results of DKL analyses with 20% random sampling measurement points as training data, detailed analyses and discussion are in the Supplementary Information. Notably, the uncertainty data of Pt/Al$_{0.93}$B$_{0.07}$N/W capacitor regarding the *linear coefficient c* parameter in Figure S2 indicates interesting structure around the high/low piezoresponse boundary.

DKL can also be implemented as an active learning method, where the DKL acquisition function is used to determine the next exploration point to search for specific physical signatures. The latter is defined by an operator-specified scalarizer function which uses the predicted response and its uncertainty as input and converts it to a measure of physical interest. For example, the scalarizer can be chosen as cubic coefficient of nonlinearity, quadratic coefficient, or maximal uncertainty of prediction. It is important to note that the pathway that the automated experiment will take depends on a chosen scalarizer, much like in human-driven experiments the specific experiment goal determines the sequence of selected measurement points. Prior to automated experiments on the microscope, DKL automated experiments were simulated with the pre-acquired data; results are shown in Figure S3 and S4.

Then, DKL-AE was used on the operational microscope. To implement DEL-AE, the Oxford Instrument Asylum Cypher was combined with in-house LabView National Instruments hardware (LabView-NI) and a Field Programmable Gate Arrays (FPGA). Figure 3 shows a schematic of the system and the data generated. The FPGA controlled the tip position by transferring the positions as electrical signals and sending them to the Asylum Cypher. LabView-NI generated BEAM waveform and acquired data after receiving a trigger from the FPGA. The DKL analyses were performed on a laptop with a GeForce RTX 3060 GPU card. The data transfer between analysis laptop and measurement computer was done with a LAN cable. First, BEPFM was performed to acquire structure images (i.e., amplitude, frequency, topography), in order to generate structure image patches used as a feature set for DKL. The first BEAM spectrum position was randomly generated by Python. Then, the DKL model analyzed the pre-defined parameter of the BEAM spectrum and the structure image patch. The DKL was trained for 200 iterations and chose the next location for measurement based on the pre-defined acquisition function, as described below. This location was conveyed to FPGA via Python code which drove the tip to the new location and triggered LabView-NI for next BEAM measurement. Then, the process was repeated. It is worth noting the novelty of this workflow compared to our previous work,[35] here real-time data transfer between two computers enables the GPU to accelerate the machine learning process.

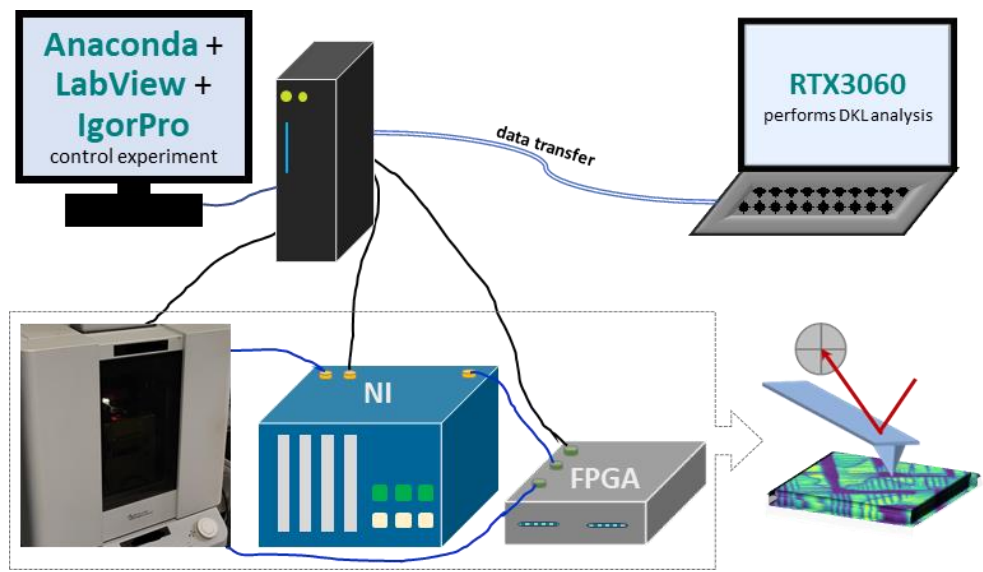

**Figure 3. A schematic of the DKL AE platform components.** The DKL AE platform includes an Oxford Instrument Asylum Cypher, an in-house LabView based National Instruments hardware (LabView-NI), a Field Programmable Gate Array (FPGA), one computer for measurement control, and one computer with a GPU to accelerate DKL analysis. In DKL AE measurements, the FPGA controls the tip via a signal to the Asylum Cypher; the LabView-NI generates the BEAM waveform and acquires data; the FPGA sends a trigger to LabView-NI for BEAM measurement after moving the tip to a desired position. Real-time data transfer between the analysis laptop and the measurement computer is enabled through a LAN cable.

PTO and $Al_{0.93}B_{0.07}N$ thin films were investigated via DKL-AE. The results are shown in Figure 4. The first row shows the topography, amplitude and frequency in a 256*256 grid, which were used to generate grid image patches for DKL. The acquisition function (AF) was derived from four parameters, cubic $a$, quadratic $b$, linear $c$ and the ratio of $a$ and $b$ ($a/b$). For both types of ferroelectric films, three structure images and four acquisition functions in DKL-AE were used, thus 24 (2*3*4) DKL-AE measurements were performed. Each DKL-AE continued for 200 steps. Figure 6 shows the DKL predictions of the corresponding parameters after 200 steps. The detailed DKL discovery processes are shown in Supplementary Information as videos of the acquisition function images with labeled exploration points.

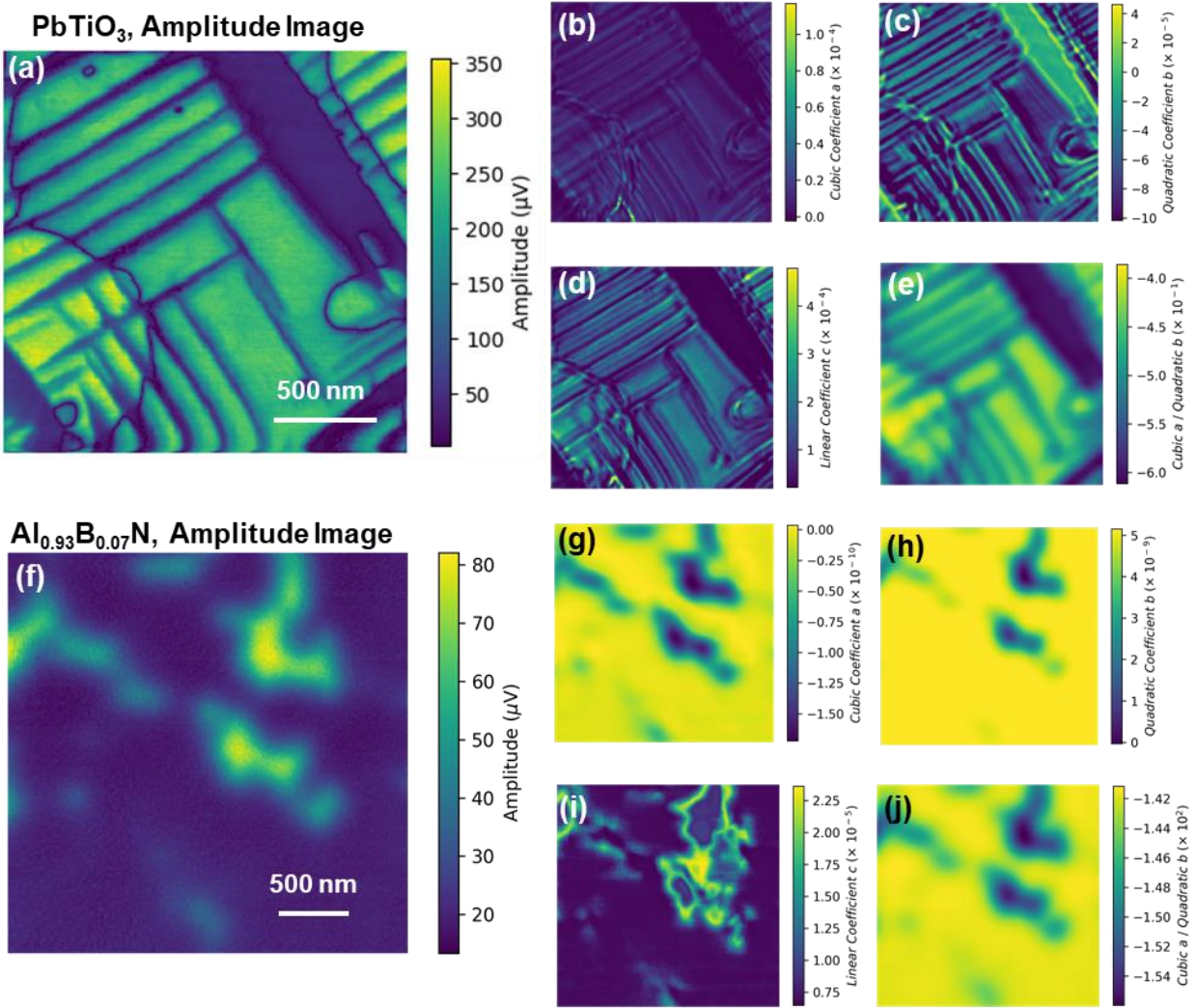

**Figure 4. DKL-BEAM results with BEPFM amplitude as the structure image.** In the DKL-BEAM measurement, a BEPFM measurement is first performed to acquire the structure images, e.g., the amplitude image. Then, four acquisition functions based on the cubic coefficient *a*, quadratic coefficient *b*, linear coefficient *c*, and the ratio of *a* and *b* were used to guide the DKL exploration. **(a)** PTO thin film BEPFM amplitude image and **(b-e)** DKL predictions according to the exploration with acquisition functions based on the cubic coefficient *a*, quadratic coefficient *b*, linear coefficient *c*, and the ratio of *a* and *b*. **(f)** the Pt/Al$_{0.93}$B$_{0.07}$N/W capacitor BEPFM amplitude image. **(g-j)** DKL predictions according to the exploration with acquisition functions based on the cubic coefficient *a*, quadratic coefficient *b*, linear coefficient *c*, and the ratio of *a* and *b*.

In the PTO thin film, the amplitude image-based DKL-AE results (Figure 4a-e) both the linear component *c* and the quadratic component *b* showed a higher response on one side and a lower response on the other side of *a/c* domain walls. In [001] tetragonal perovskite films such as PbTiO$_3$, the *a/c* domain walls generally lie along [101] planes, and so are tilted at 45° with respect to the sample surface. The amplitude contrast across the domain wall is thus a function of the volume of material probed and the number of *a*- and *c*-domains in that volume. Near *c/c* domain walls there is a larger quadratic component *b* (Figure 4b(II)); this region is slightly broader than the image of the domain walls themselves. Topography and frequency image-based DKL-AE results are shown in Figure S5, which indicated features related to surface geometry and *a/c* domains.

In a Pt/Al$_{0.93}$B$_{0.07}$N/W capacitor, both topography and frequency (Figure S6) based results shown no obvious spatial variation likely because topography and frequency responses originate from the top electrode Pt layer while the piezoresponse originates from the Al$_{0.93}$B$_{0.07}$N layer. The results based on amplitude (Figure 4f-j) indicate lower *cubic a* and *quadratic b* in the high piezoresponse domains, and intermediate *cubic a* and *quadratic b* in the high/low piezoresponse boundary. The displacement – voltage curves were very slightly concave down. Most interestingly, the *linear c* shows a sharp contrast across the high/low piezoresponse boundary—the *linear coefficient c* is higher at the low piezoresponse side but lower at the higher piezoresponse side. A hypothesis that would be consistent with the observations for the Al$_{0.93}$B$_{0.07}$N layer is the appearance of wedge domains having opposite polarity of the matrix, which lead to a lower volume thickness-averaged piezoelectric response, but a higher activity level under a large signal drive field.

Figure 5 shows a comparison of exploration points when different structure images were used. Notably, most exploration points locate near *a/c* domain walls in the PTO thin film when the acquisition function is the quadratic coefficient *b* (Figure 5b) regardless of what structure image is used. In addition, most exploration points are located near *c/c* domain walls in the PTO film when the acquisition function is the linear coefficient *c* (Figure 5c) in all cases. Figure 5e-h shows Pt/Al$_{0.93}$B$_{0.07}$N/W results. Exploration points based on topography and frequency were mostly scattered, while the exploration points based on amplitude were mainly located around high/low piezoresponse domain boundaries.

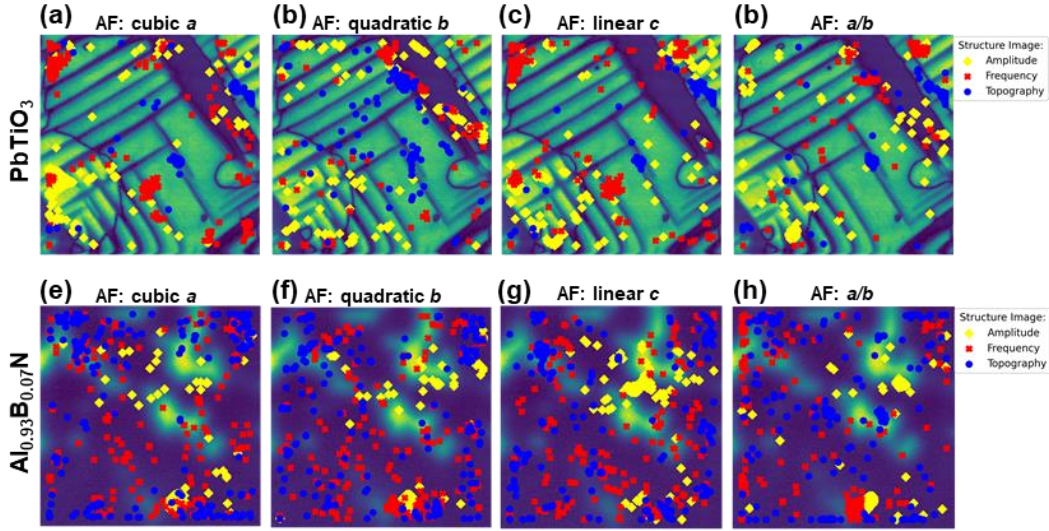

**Figure 5. Experimental DKL navigated exploration points in DKL-BEAM measurements. (a-d),** exploration points of PTO thin film results with acquisition functions as cubic *a*, quadratic *b*, linear *c*, and *a/b*, respectively. **(e-h)**, exploration points in Pt/Al$_{0.93}$B$_{0.07}$N/W capacitor results with acquisition functions as cubic *a*, quadratic *b*, linear *c*, and *a/b*, respectively. The corresponding structure images used in the explorations are indicated via colors.

**Conclusions**

Automated band excitation variational amplitude microscopy (BEAM) based on deep kernel learning (DKL) was implemented to explore the local nonlinear electromechanical response in a model ferroelectric PTO film and a Pt/Al$_{0.93}$B$_{0.07}$N/W capacitor. DKL-BEAM explorations suggested asymmetric nonlinear behavior across *a/c* domain walls and a broadened high nonlinear response region around *c/c* domain walls in the model PTO film. DKL-BEAM explorations indicate that the highest linear piezoelectric response was coupled with the lowest nonlinear response in the Pt/Al$_{0.93}$B$_{0.07}$N/W capacitor. Formulating dissimilar exploration strategies in deep kernel learning as alternative hypotheses established the preponderant physical mechanisms behind non-linear behaviors. Automated experiments show potential in multimodal imaging to discern competing physical mechanisms and can be used in electron, probe, and chemical imaging.

**Methods**

*Data analysis*

The DKL methodologies are established in Jupyter notebooks and are available from https://github.com/yongtaoliu/DKL-Nonlinearity.

*PTO sample*

The PTO film was grown on a KTaO$_3$ substrate with a SrRuO$_3$ bottom electrode by chemical vapor deposition

*AlBN sample*

Pt/Al$_{0.93}$B$_{0.07}$N/W film was grown on (001) Sapphire substrate by using pulsed DC sputtering, with (110) W bottom and Pt top electrode that both deposited by DC magnetron sputtering.

*BEPFM and BEAM measurements*

The PFM was performed using an Oxford Instrument Asylum Research Cypher microscope with Budget Sensor Multi75E-G Cr/Pt coated AFM probes (~3 N/m). Band excitation data are acquired with a National Instruments DAQ card and operated with a LabView framework.

**Conflict of Interest**

The authors declare no conflict of interest.

**Authors Contribution**

S.V.K. conceived the project and M.Z. realized the DKL workflows. Y.L. performed detailed analyses. Y.L. deployed the DKL to BEAM measurement and obtained results. K.K. made a script enabling communication between FPGA and LabView. H.F. provided the PTO sample. J.H. and J.P.M. provided the Al$_{0.93}$B$_{0.07}$N sample. W.Z. prepared samples with different levels of wake-up. S.T.M. is responsible for the interpretation of the nonlinearity responses. All authors contributed to discussions and the final manuscript.

**Acknowledgements**

This effort (wurtzite film growth, SPM, data analysis) was supported as part of the center for 3D Ferroelectric Microelectronics (3DFeM), an Energy Frontier Research Center funded by the U.S. Department of Energy (DOE), Office of Science, Basic Energy Sciences under Award Number DE-SC0021118, and the (workflow development) Oak Ridge National Laboratory's Center for Nanophase Materials Sciences (CNMS), a U.S. Department of Energy, Office of Science User Facility.

**Data Availability Statement**

The data that support the findings of this study are available at https://github.com/yongtaoliu/DKL-Nonlinearity.